\documentstyle[11pt]{article}
\newcommand{\beq}{\begin{equation}}
\newcommand{\eeq}{\end{equation}} 

\newcommand{\beqa}{\begin{eqnarray}}
\newcommand{\eeqa}{\end{eqnarray}}

\def\half{\frac{1}{2}}
\def\opone{\leavevmode\hbox{\small1\kern-3.8pt\normalsize1}}

\def\A{{\cal A}}
\def\B{{\cal B}}
\def\sign{{\rm sign}}
 
\begin{document}

\title{A local hidden variable model of quantum correlation exploiting the detection loophole}
\author
{N. Gisin and B. Gisin\\
\protect\small\em Group of Applied Physics, University of Geneva, 
1211 Geneva 4, Switzerland}
\date{\today}

\maketitle

\begin{abstract}
A local hidden variable model exploiting the detection loophole
to reproduce exactly the quantum correlation of the singlet
state is presented. The model
is shown to be compatible with both the CHSH and the CH Bell inequalities. 
Moreover, it bears the same rotational symmetry as spins. The reason why the
model can reproduce the quantum correlation without violating the Bell theorem is
that in the model the efficiency of the detectors depends on the local hidden
variable. On average the detector efficiency is limited to 75\%.
\end{abstract}

\section{Introduction}\label{introduction}
Quantum theory is nonlocal in the sense that it predicts that distant systems can
produce outcomes 
whose correlation can not be explained by any model based only on local variables
(i.e. in which results "here" happen independently of choices made at a distance),
as demonstrated by Bell \cite{Bell64}.
The state of such "quantum correlated" systems is said to be "entangled".
Because these outcomes are stochastic, there is no way to use them for signaling. Hence,
despite this nonlocality, there
is no direct conflict with relativity \cite{Popescu}.

All experiments to date are in perfect agreement with
quantum theory. There is thus strong evidence for the view that Nature is nonlocal. 
However, considering the importance of such a conclusion, the experimental evidence should be
analyzed very carefully and any additional experimental tests of quantum entanglement
are most welcome.

There is one logical loophole affecting all the experiments carried out so far that
deserves special attention. This is the "detection loophole" \cite{Pearle70,detloophole}, 
which is the central issue of 
this letter. The detection loophole is based on the following fact: in real experiments
the efficiencies of the detectors are such that the set of detected
events
is significantly smaller than the set of tested quantum systems. One has thus to assume
that the sample over which the statistics is measured is a fair sample. From the point of
view of quantum mechanics this assumption is almost trivial. However, from the 
local hidden variable ({\bf lhv})
point of view the opposite assumption is equally almost trivial: if there are additional
variables, unknown today, it is very plausible that the actual value of these variables
affects the probability to trigger a detector.

Admittedly, invoking the detection loophole to explain all the experimental tests of Bell
inequality is somewhat artificial. However, it is annoying that this loophole resists after
almost 30 years \cite{Pearle70} of research and progress! In this letter we present a simple 
lhv model which reproduces analytically the quantum correlations corresponding to the
singlet state of two spin $\half$:
\beq
E(\vec a,\vec b)=-\vec a\vec b
\label{Eab}
\eeq
where $\vec a$ and $\vec b$ represent the bases of the two distant measurements. The quantum
correlation of any other maximally entangled states can also be reproduced by our 
model by applying to the lhv (which are vectors) the same local rotations that 
transform any maximally entangled state into the singlet. However, the model is 
compatible with all Bell inequalities. In our model the detector
efficiency is of 75\%. This is much larger than the efficiency used in all Bell tests so far, 
hence our model reproduces all existing experimental data of Bell tests.

In the next section we present our lhv model. 
Then, in section 3 we establish the connection with the Clauser-Horne (CH) Bell 
inequality \cite{CH74}. In the conclusion we comment on the connection between our lhv
model and the paper and the paper by Steiner \cite{Steiner99} in which the problem of minimal
classical communication for simulating quantum correlation is considered and which inspired
our work very much.

\section{A local hidden variable model}
In this section we present our lhv model, inspired by Steiner's contribution \cite{Steiner99}.
We first present it in its simplest form, which is asymmetric between Alice and Bob, the
two observers. Next we show how to make
it symmetric. In this model, each spin $\half$ is characterized, in addition to its quantum state
$\rho$, by a normalized classical arrow $\vec\lambda$ with uniform a priori probability distribution. 
If the spin is measured along a direction $\vec a$, the outcome $\pm1$
is determined by the sign of the scalar product $(<\vec\sigma>_\rho-\vec\lambda)\cdot\vec a$, 
where $<\vec\sigma>_\rho=Tr(\vec\sigma\rho)$ 
denotes the expectation values of the Pauli matrices. For a single
spin $\half$, the probability of an outcome $+1$ is then
in accordance with the quantum prediction for a state $\rho$:
$P_+(\vec a)=\half(1+<\vec\sigma>_\rho\cdot\vec a)$. In case of two spin $\half$ of total spin zero (i.e.
singlet state), the vectors characterizing each spin are opposite: $\vec\lambda_\B
=-\vec\lambda_\A$. With these rules the correlation is linear (see Fig. 1) \cite{Belllhv}: 
\beq
E_{lhv}(\vec a,\vec b)=-1+\frac{2\theta_{ab}}{\pi}
\label{Elhv}
\eeq
where $\theta_{ab}\in[0..\pi]$ is the
angle between $\vec a$ and $\vec b$. They do not violate the CHSH-Bell inequality.
So far measurements always produce outcomes, i.e. we
did not exploit the detection loophole. Now, let's assume that whenever a measurement
in direction $\vec a$ is performed, an outcome on Alice detector 
is produced only with probability 
$|\vec\lambda_\A\cdot\vec a|$. That is, on Alice side no outcome at all is produced in a ratio
$1-|\vec\lambda_\A\cdot\vec a|$ of cases, while Bob always produces an outcome. 
This affects the correlation; indeed, when
$\vec a$ happens to be close to $\vec\lambda_\A$ then the probability that an outcome is
produced is larger than when $\vec a$ happens to be nearly orthogonal to $\vec\lambda_\A$.
To compute the correlation function $E(\vec a,\vec b)$, we first need the mean probability $p$ 
that an outcome is produced:
\beq
p=\int_{S^2}\frac{d\vec\lambda}{4\pi}|\vec a\cdot\vec\lambda| =\half
\eeq
Next, we need the conditional density probability distribution of the $\vec\lambda_\A$ given that
an outcome is produced:   
\beqa
\rho(\vec\lambda~|~{\rm outcome~ produced})&=&
\frac{\rho(\vec\lambda ~{\rm and~ outcome~produced})}{{\rm Prob}({\rm outcome~produced})} \\
&=&\frac{\frac{1}{4\pi}|\vec a\cdot\vec\lambda|}{p} \\
&=&\frac{1}{2\pi}|\vec a\cdot\vec\lambda|
\eeqa
Now, the correlation can be computed (note that it is convenient to chose a reference
frame such that $\vec b=(0,0,1)$, $\vec a=(\sin(\alpha),0,\cos(\alpha))$ and
$\vec\lambda=(\sqrt{1-\eta^2}\cos(\phi),\sqrt{1-\eta^2}\sin(\phi),\eta)$, $d\vec\lambda=
d\eta d\phi$):
\beqa
E(\vec a,\vec b)&=&\int_{S^2}d\vec\lambda~\rho(\vec\lambda~|~{\rm outcome~produced})~
\sign(\vec a\cdot\vec\lambda)~\sign(-\vec b\cdot\vec\lambda) \\
&=&-\int_{S^2}d\vec\lambda~\frac{\vec a\cdot\vec\lambda}{2\pi}~\sign(\vec b\cdot\vec\lambda) \\
&=&-\int_{-1}^1d\eta \int_0^{2\pi}\frac{d\phi}{2\pi}\left(\eta\cos(\alpha)
+\sqrt{1-\eta^2}\cos(\phi)\sin(\alpha)\right)\sign({\eta}) \\
&=&-\cos(\alpha)=-\vec a\cdot\vec b
\eeqa
Witch agrees exactly with the quantum correlation (\ref{Eab}).

In summary, if Alice's detector is allowed to fire only half the time, with a probability
to fire that depends on the lhv $\lambda_\A$,
while Bob's detector always fires, then the quantum
correlation can be recovered exactly! The asymmetric protocol presented above can clearly
be made symmetric: it suffices to add an additional binary local hidden variable which determines
when Alice and Bob exchange their roles. If this binary variable is randomly distributed,
then both Alice's and Bob's detector will fires in 75\% of cases, that is both detectors
show an effective efficiency of 75\%, much above the efficiencies of the detectors used in 
all actual experimental Bell tests. Note that in the model presented so far there is always
at least one detector that fires. This may be considered as bizard and corrected for by
adding to the model yet another binary local hidden variable which programs both
spins to produce no outcome at all. If this "no outcome at all" happens with probability
1/9, then both detectors appear to be independent, each with an efficiency of $\frac{8}{9}
\frac{3}{4}=\frac{2}{3}$:
probability of 2 outcomes = $(\frac{2}{3})^2$, of 1 outcome = $\frac{2}{3}\frac{1}{3}$ and
of 0 outcome = $(\frac{1}{3})^2$.

A simple software demo that simulates Alice and Bob "experiments" is available on our
WEB side \cite{soft}. It allows two independent computers to produce data which violate
the Bell inequality.

\section{Connection to the Clauser-Horne inequality}
The most wellknown Bell inequality is the CHSH one \cite{CHSH69}:
\beq
E(\vec a,\vec b)+E(\vec a,\vec b')+E(\vec a',\vec b)-E(\vec a',\vec b') \le 2
\label{CHSH}
\eeq
where $E(\vec a,\vec b)=P_{++}(\vec a,\vec b)+P_{--}(\vec a,\vec b)
-P_{+-}(\vec a,\vec b)-P_{-+}(\vec a,\vec b)$ is the correlation function with
$P_{ij}(\vec a,\vec b)$ the probabilities of outcomes $ij=\pm\pm$ when the measurement
bases are defined by the directions $\vec a$ and $\vec b$. In experiments one does not
measure probabilities, but coincidence rate (i.e. coincidence counts per time unit)
$N_{ij}(\vec a,\vec b)$. The correlation function $E(\vec a,\vec b)$ is thus evaluated
experimentally as a "renormalized correlation":
\beq
E(\vec a,\vec b)=\frac{N_{++}+N_{--}-N_{+-}-N_{-+}}{N_{++}+N_{--}+N_{+-}+N_{-+}}
\eeq
where for compactness we have dropped the indication of the bases. This renormalization,
although natural in the frame of quantum mechanics, is questionable. Indeed, with this
renormalization the lhv model presented in the previous section, although
purely local, does violate
the CHSH inequality (\ref{CHSH}). Hence, in order to definitively prove all lhv models
wrong, the experimental data should violate an inequality involving no "renormalization",
ideally involving only count rates.

The CHSH inequality (\ref{CHSH}) can be deduced from the following trivial one:
\beq
ab+ab'+a'b-a'b' = a(b+b')+a'(b-b')\le2
\eeq
where $a,a',b,b'\in\{-1,+1\}$. In order to obtain an inequality between count rates, we
introduce numbers $x=\half(1+a)$ and $y=\half(1+b)$, $x,y\in\{0,1\}$ such that:
\beq
ab+ab'+a'b-a'b'-2=4\left(xy+xy'+x'y-x'y'-(x+y)\right)\le0
\label{xy}
\eeq
From the inequality (\ref{xy}), one deduces the CH inequality \cite{CH74}:
\beq
N_{++}(\vec a,\vec b)+N_{++}(\vec a,\vec b')+N_{++}(\vec a',\vec b)-
N_{++}(\vec a',\vec b') \le N_{+\cdot}(\vec a)+N_{\cdot+}(\vec b)
\label{CH}
\eeq
where $N_{+\cdot}(\vec a)$ and $N_{\cdot+}(\vec b)$ denote the single counts on the 
first and second system, respectively.

Quantum mechanics predicts:
\beqa
N_{+\cdot}(\vec a)=P_{+\cdot}(\vec a)\eta_{_\A} N \hspace{1cm}
N_{\cdot+}(\vec b)=P_{\cdot+}(\vec b)\eta_{_\B} N \label{Ndot} \\
N_{++}(\vec a,\vec b)=P_{++}(\vec a,\vec b)\eta_{_\A}\eta_{_\B} N
\eeqa
where $N$ is the total number of quantum systems tested per unit time and 
$\eta_\A$ and $\eta_\B$ denote the efficiencies of the first and second detector,
respectively \cite{footnote1}. Accordingly, assuming for simplicity 
$\eta_\A=\eta_\B\equiv\eta$, quantum mechanics predicts a violation of the CH inequality
(\ref{CH}) if and only if:
\beq
\eta>\frac{P_{+\cdot}(\vec a)+P_{\cdot+}(\vec b)}
{P_{++}(\vec a,\vec b)+P_{++}(\vec a,\vec b')+P_{++}(\vec a',\vec b)-P_{++}(\vec a',\vec b')}
\eeq
The optimal is obtained for the same setting $\vec a,\vec a',\vec b,\vec b'$ that
maximize the violation of the CHSH inequality (\ref{CHSH}). For these one
obtains $\eta_{\rm threshold}=\frac{2}{1+\sqrt{2}}\approx 0.828$. Hence, quantum mechanics
predicts that the CH inequality (\ref{CH}) can be violate, and thus lhv models 
definitively ruled out, only with detector's efficiencies higher than 82.8\%! This is
in agreement with our model in which the "efficiency" is of "only" 75\%. Note however
that in our model the detection probabilities on both sides are not independent:
$\overline{\eta_{_\A}\eta_{_\B}}\ne\bar\eta_{_\A}\cdot\bar\eta_{_\B}$. Hence, the
"relevant efficiency" is even smaller, in full agreement with the model with independent
detectors discussed at the end of the previous section:
\beq
\frac{\overline{\eta_{_\A}\eta_{_\B}}}{\bar\eta_{_\A}}=\frac{p}{\half(p+1)}
=\frac{2}{3}<\eta_{\rm threshold}
\label{etaS2_2}
\eeq

We conclude this section with some comments. First, it could be that some other inequality
involving only count rates could be violated with detector efficiencies bellow 82.8\%.
However, our model proves that no such inequality could achieve a threshold bellow 75\%
(actually, no such inequality is known to us). Next, in our model there is always at
least one detector that fires. This may seem unphysical. But this is a point that has
never been ruled out experimentally. Moreover if one allows for lower detector efficiencies
one could easily generalize our model and introduce cases where no detector at all fire.
Finally, the optimal directions $\vec a,\vec a',\vec b,\vec b'$ lie in a plane, hence
a model reproducing the quantum correlation for settings restricted to a plane is also of
interest. Such a model follows straightforwardly from Steiner contribution \cite{Steiner99},
with a detection efficiency of $\half(1+\frac{2}{\pi})\approx 81.8\%$ and a "relevant
efficiency" (taking into account the correlation between the detection probabilities) of
$77.8\%$.

\section{Conclusion}\label{conc}
The detection loophole remains open after almost 30 years of research and progress,
despite its importance for the understanding of physics and for applications in
quantum communication. The model presented in section 2 underlines how simple and
natural a local model can be, while reproducing exactly the quantum correlation function
thanks to this loophole. This model also proves that there is no hope to close the
detection loophole with detectors' efficiencies bellow 75\%.

Our lhv model was inspired by the CH model \cite{CH74} and by Steiner's contribution
\cite{Steiner99}. In the latter an apparently different problem was considered, namely
that of the classical communication cost of simulating quantum correlation. The idea is
that Alice and Bob have common classical information and exchange as little classical
information as possible to reproduce the correlation (\ref{Eab}), thus violating
Bell inequality. Clearly, any lhv model
provides a classical communication simulation model: instead of producing no outcome,
Alice informes Bob not to use their common information $\vec\lambda_j$ and to go
to the next one $\vec\lambda_{j+1}$. The detection efficiency translates then into
the communication cost. Conversely, a classical communication simulation model provides
a lhv model. But in this direction the connection is not straightforward. Indeed, Bob's
action could depend on the information he receives from Alice in a more complicated way
than merely going to the next common information.
Nervertheless, a lhv obtains if Alice and Bob decide in advance to bet on the action Bob
would have to take on each instance and that Alice merely produces no outcome whenever
the actual action Bob should carry out differs from the bet. This can be very inefficient,
but provides a lhv model.

The local symmetry of our model is the same rotational symmetry that characterizes spins.
However, our model does not reproduce "coherent measurements", like the famous Bell
measurements so powerful in quantum information processing, in particular for
entanglement swapping \cite{EntSwapTh}. Nor does it account for
partially entangled states (which have surprising characteristics, see \cite{Eberhard93}), 
nor for generalized (POVM) measurements. 
But, to date, neither Bell measurements nor tests of Bell
inequality using partial entanglement have been performed. This work might provide
additional motivation for such experiments.

\small
\section*{Acknowledgments}
This work profited from enjoyable and stimulating discussions with Sandu Popescu.
It was partially supported by the Swiss National Science Foundation.

\section*{Figure caption}
Correlation functions, with $\theta_{ab}$ the angle between the measurement directions
on Alice and Bob sides. The dotted line corresponds to the local hidden variable model
with 100\% detection efficiency (\ref{Elhv}), the full line to our model 
with 75\% detection efficiency and to quantum mechanics (\ref{Eab}).

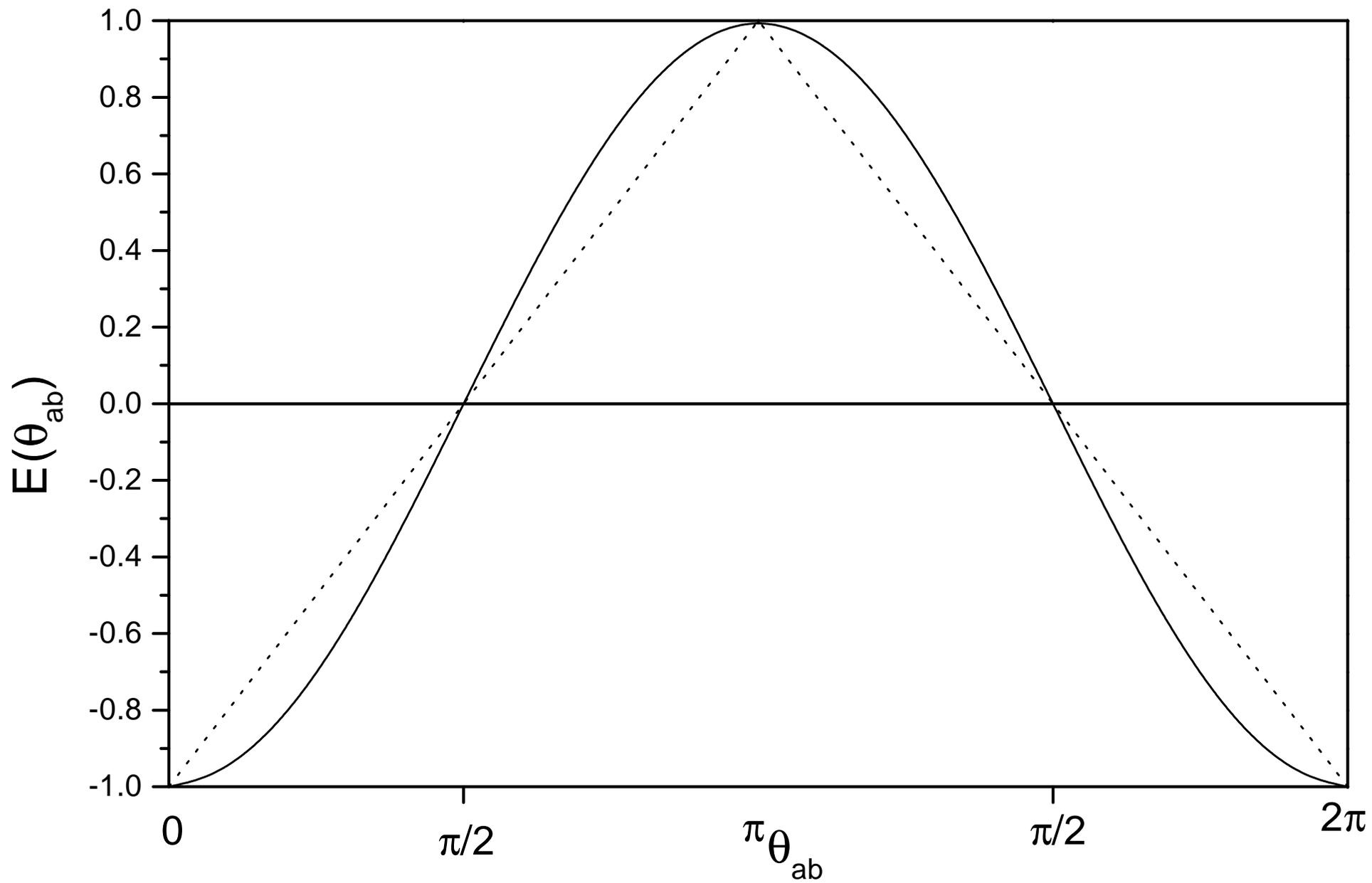
\end{document}